\documentclass[twocolumn,aps,prb,amsfonts,amssymb,amsmath,showpacs,english]{
revtex4-1}
\usepackage[T1]{fontenc}
\usepackage[latin9]{inputenc}
\usepackage{amssymb}
\usepackage{esint}
\usepackage{babel}
\usepackage{graphicx}

\begin{document}

\title{\bf Impurity effects and ferromagnetism in excitonic insulators}
\author{Jian Li, Ning Hao and Yupeng Wang}
\affiliation{Beijing National Laboratory for Condensed Matter
Physics, Institute of Physics, Chinese Academy of Sciences,
Beijing 100190, PR China}

\begin{abstract}
Both nonmagnetic and magnetic impurity effects in spin singlet and triplet
excitonic insulators were investigated.
The bound state energies caused by single impurity were given. The different compositions of the bound states can be used to detect the symmetry of the excitonic insulators. In
finite concentration problems, nonmagnetic impurities showed same pair-breaking
effect in singlet and triplet excitonic insulators while magnetic impurities
showed weaker pair-breaking effect in triplet excitonic insulators than in singlet ones. The pair-breaking effects suppressed the ferromagnetic range via doping and gave a
natural explaination for experimental results.
\end{abstract}


\maketitle
\begin {section}{Introduction}
The excitonic phase has been researched for a long time since
1960'\cite{J,Rice1,YuV}.
It was proposed in semimetals, in which there is a band overlap and
the normal states are unstable. The true ground state becomes a state
with electron-hole pairs due to the Coulomb interaction. This kind
of state can also occur in semiconductors once the energy gap
becomes smaller than the binding energy of an exciton\cite{J}. This so-called
'excitonic insulator' has been discussed for many years theoretically but
materials in experiment were seldom reported so far. $1T-TiSe_2$, among the
transition-metal dichalcogenides,  may be one example of the excitonic
insulators which obtained more and more direct evidences from
experiments\cite{THP,HCE}. Another possible example is CaB$_6$. The latter one
has gained more attention since weak ferromagnetism
was found in slightly doped CaB$_6$ in 1999\cite{Y}. Theoretically many
authors relate this to the
historical excitonic insulators\cite{Rice,Gorkov,Varma}. By now there are still
debates on this proposal, especially about the band gap. The angle-resolved photoemission
spectrum showed a gap of 1ev\cite{Maiti,JD} which is too large for excitonic instability and another large gap of 0.8ev
is given by the LDA+GW calculations\cite{HJ}. Band overlap and semimetal
framework were demonstrated by the de Haas-van Alphen and Shubnikov-de Haas
experiments\cite{RG,MC,DH} and local density approximation band structure
calculations\cite{AH,CO,HK}.
Though there are controversies on the band structure, the excitonic insulator
scenario is nevertheless a very promising explaination for the main
physics in CaB$_6$.

In Ca$_{1-x}$La$_{x}$B$_{6}$, slightly La doping would induce not only the carriers but also impurities/disorders
inevitably and the impurity effects played an important role in
the formation of the weak ferromagnetism. According to the first principle
calculations, the defects or vacancies in the boron sublattice can result in
impurity bands or localized gap states and this modification may be important
to the origin of ferromagnetism in CaB$_6$\cite{Maiti2,JX}. From the language
of
excitonic insulators, the impurities will induce in-gap bound states and
it was confirmed by tunneling measurements on high-quality defect-controlled
single crystals\cite{BK}. Therefore the impurity effect can not be neglected
in the excitonic insulator theory for CaB$_6$. To explore how the charged
impurity doping can affect the excitonic insulators constitutes one of the
motivations
of this paper. Another motivation originates from the incompleteness of the research on
impurity effects in excitonic insulators. Historically the impurity
scattering in excitonic insulators was investigated by J. Zittartz in
1967\cite{JZ} and in that work only intra-band nonmagnetic impurity in exciton
insulators with no regard for the spin freedom was studied. Inter-band
nonmagnetic
impurity scattering was investigated by T.Ichinomiya\cite{TI} and other works
concentrated in the bilayer systems\cite{Dubi,RB} in which the spin freedom was
also neglected.
In this paper we will discuss and compare the inter-/intra-band
magnetic/nonmagnetic
impurity effect in singlet and triplet excitonic insulators. We will also modify the exciton
insulator scenario on ferromagnetism in Ca$_{1-x}$La$_{x}$B$_{6}$.

This paper is constructed as follows: In the first section we will introduce
our model and Hamiltonian. Then we will give the locations of
the bound states induced by single magnetic or nonmagnetic impurity.
Finite concentration problems of impurities and pair breaking effects
will be discussed in the third section. In the fourth section we will
investigate
the impurity caused-modification to the ferromagnetism in doped exciton
insulators. We will leave the summary and further discussion to the last
section.
\end {section}

\begin {section}{Model and Hamiltonian}
We introduce the noninteracting Hamiltonian in a two band system:
\begin{eqnarray*}
H_{0}=\sum_{k,\sigma}(\varepsilon_{k}^{a}a_{k,\sigma}^{\dagger}a_{k,\sigma}
+\varepsilon_{k}^{b}b_{k,\sigma}^{\dagger}b_{k,\sigma})
\end{eqnarray*}
here the kinetic energy in two electron bands are
$\varepsilon_{k}^{a}=\frac{\hbar^{2}k^{2}}{2m_{a}}-\frac{1}{2}E_{G}$,
$\varepsilon_{k}^{b}=-\frac{\hbar^{2}k^{2}}{2m_{b}}+\frac{1}{2}E_{G}$,
in which $E_{G}$ is the band gap or overlap and $|E_G|$ should be quite small to
keep the ground state remain in the excitonic phase.  Without losing generality, we
consider
two symmetric bands, which means $m_{a}=m_{b}=m$. Now we introduce
the interaction term, which reads:
\begin{eqnarray*}
H_{1}=\frac{U_{1}}{\Omega}\sum_{k,k',q,\sigma,\sigma'}a_{k+q,\sigma}^{\dagger}b_
{k'-q,\sigma'}^{\dagger}b_{k',\sigma'}a_{k,\sigma}
\end{eqnarray*}
where $U_{1}$ represents the interaction term which comes from the Coulomb
interaction. In most of the
bulk samples, the system will choose singlet or triplet excitonic phase
as the ground state\cite{RMP} because there is a finite splitting between singlet and triplet
states. The splitting may come from short-range Coulomb terms which favor triplet state, or 
electron-phonon interaction which favors singlet state\cite{Rice3}. 
In the presence of nesting effect, if the screened Coulomb
interaction term dominates there will be a degeneracy
between singlet and triplet excitonic phase. In this case there is an
SU(2)$\times$SU(2)
symmetry in the excitonic insulators. In bilayer systems this splitting is
exponentially
small because of the inter-layer distance and the barriers between
two layers\cite{Shim}, and spin freedom can thus be neglected. It is also believed that the splitting in CaB$_6$ is quite small\cite{Gorkov}.

Now we introduce the singlet and triplet order parameters: $\Delta_{s}$
and $\Delta_{T}$ and rewrite the mean-field Model into Nambu space:
By introducing the vector
$\psi=(a_{k\uparrow},a_{k\downarrow},b_{k\uparrow},b_{k\downarrow})$,
the Hamiltonian can be written as :
\begin{eqnarray*}
H_{MF}=\sum_{k}\psi^{\dagger}M\psi
\end{eqnarray*}
where M is a $4\times4$ matrix, then the bare Green function for
singlet excitonic insulators is
\begin{eqnarray}
&&G_{0}^{-1}(k,\omega)=\omega-H_{MF}^S\nonumber\\
 &&=\left(\begin{array}{cc}
(\omega-\frac{\hbar^{2}}{2m}k^{2}+\frac{E_{g}}{2})\cdot I & -\Delta_{S}\cdot I\\
-\Delta_{S}\cdot I & (\omega+\frac{\hbar^{2}}{2m}k^{2}-\frac{E_{g}}{2})\cdot
I\end{array}\right)
\end{eqnarray}
and triplet excitonic insulators:
\begin{eqnarray}
&&G_{0}^{-1}(k,\omega)=\omega-H_{MF}^T\nonumber\\
&&=\left(\begin{array}{cc}
(\omega-\frac{\hbar^{2}}{2m}k^{2}+\frac{E_{g}}{2})\cdot I &
-\Delta_{T}\cdot\sigma\\
-\Delta_{T}\cdot\sigma & (\omega+\frac{\hbar^{2}}{2m}k^{2}-\frac{E_{g}}{2})\cdot
I\end{array}\right)
\end{eqnarray}

The impurity scattering term can be written as:
\begin{eqnarray*}
V=\left(\begin{array}{cc}
J_{1n}\cdot I+J_{1m}\sigma\cdot S & J_{2n}+J_{2m}\sigma\cdot S\\
J_{2n}+J_{2m}\sigma\cdot S & J_{1n}\cdot I+J_{1m}\sigma\cdot
S\end{array}\right)\
\end{eqnarray*}
in which $\nu=n,m$ represents the nonmagnetic and magnetic impurity 
scattering. S denotes the spin operator of the magnetic impurity and $i=1,2$ represents the intra-band and inter-band channel, respectively.
\end {section}

\begin {section}{Single impurity problems}
To obtain the bound state energy, one has to get the T-matrix which
can be written as:
\begin{eqnarray*}
T=\frac{V}{1-G_{0}(r=0,\omega)V}
\end{eqnarray*}
the pole of the T-matrix gives the bound state energy. In singlet
excitonic insulators, for nonmagnetic impurity,
\begin{eqnarray*}
&&E_{B1}=\frac{\Delta_{S}[1-(\alpha_{1n}+\alpha_{2n})^{2}]}{1+(\alpha_{1n}
+\alpha_{2n})^{2}};\ \ \ \ \ \ \rm{\Delta_S<0}\\
&&E_{B2}=\frac{-\Delta_{S}[1-(\alpha_{1n}-\alpha_{2n})^{2}]}{1+(\alpha_{1n}
-\alpha_{2n})^{2}};\ \ \ \ \rm{\Delta_S(\alpha_{2n}-\alpha_{1n})<0}
\end{eqnarray*}
where $\alpha_{1n}=J_{1n}\pi N_{F}$ , $\alpha_{2n}=J_{2n}\pi N_{F}$.
And for magnetic impurity case,
\begin{eqnarray*}
&&E_{B1}=\frac{\Delta_{S}[1-(\alpha_{1m}+\alpha_{2m})^{2}]}{1+(\alpha_{1m}
+\alpha_{2m})^{2}}\\
&&E_{B2}=\frac{-\Delta_{S}[1-(\alpha_{1m}-\alpha_{2m})^{2}]}{1+(\alpha_{1m}
-\alpha_{2m})^{2}}
\end{eqnarray*}
where $\alpha_{1m}=J_{1m}\pi N_{F}S(S+1)$ , $\alpha_{2m}=J_{2m}\pi N_{F}S(S+1)$.
Notice some extreme cases in the singlet excitonic insulators, for the negative
order parameter and if there
is only one scattering channel exists, the bound state energy will
have the same form of the single magnetic impurity in s-wave
superconductors\cite{Yu,Shiba,Rusinov},
which reads $E_{B}=\frac{\pm\Delta(1-\alpha_{i}^{2})}{1+\alpha_{i}^{2}}$
(i=1n,2n,1m,2m). In this case the bound state energy will move to
the Fermi energy with the increasing of the scattering constant while move
to the gap-edge side with the decreasing of impurity scattering. If there is no
impurity scattering ($J_{i\nu}=0$), the excitation will emerge into
the normal excitonic gap $E_{B}=\pm\Delta$.

In triplet case, the bound state energy can be obtained in the same way. For
single nonmagnetic impurity
\begin{eqnarray*}
E_{B1}=\pm\frac{\Delta_{T}[1-(\alpha_{1n}+\alpha_{2n})^{2}]}{1+(\alpha_{1n}
+\alpha_{2n}^{2}};\ \ \ \ \rm{\pm(\alpha_{1n}+\alpha_{2n})\Delta_T<0}\\
E_{B2}=\pm\frac{\Delta_{T}[1-(\alpha_{1n}-\alpha_{2n})^{2}]}{1+(\alpha_{1n}
-\alpha_{2n})^{2}};\ \ \ \ \rm{\pm(\alpha_{1n}-\alpha_{2n})\Delta_T<0}
\end{eqnarray*}
Since the form of the bound state energy for single magnetic impurity in triplet
excitonic insulators is quite complex, we only give the numerical results.
The bound state energies for single nonmagnetic and magnetic impurity in singlet and
triplet excitonic insulators are shown in Fig1.
 \begin{figure}[tbh]
\label{Fig1}\includegraphics[width=3.6in] {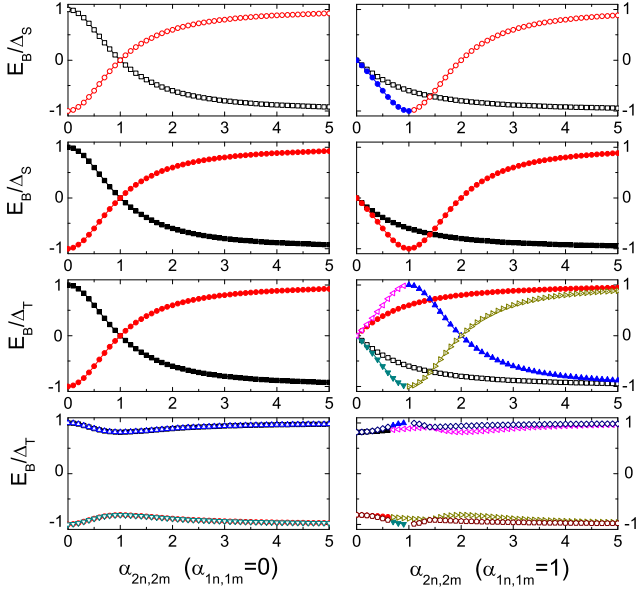}
\caption{Bound state energy induced by single impurity in singlet
and triplet excitonic insulators via the scattering strength. From top to below
is the case for nonmagnetic/magnetic impurity in singlet excitonic insulators,
nonmagnetic/magnetic impurity in triplet excitonic insulators, respectively.
Please note that in the figure solid symbols represent $\Delta_{S,T}>0$ and open
symbols represent $\Delta_{S,T}<0$.}
\end{figure}

From Fig1 we can see that location of the bound state energy which is induced by
single impurity is very different between
singlet excitonic insulators and triplet ones. Furthermore,
in excitonic insulators, the quasiparticle
induced by single nonmagnetic or magnetic impurity can consist one or two hole
($\omega<0$) or one or two particle type ($\omega>0$) states in some range of
$J_{1n}$, $J_{2n}$ or $J_{1m}$, $J_{2m}$. This is quite different from the
single magnetic impurity in s-wave superconductors in which hole and electron
type excitations must appear symmetrically at the same time. It is well known that
 both the hole- and electron-type components of the bound states can be detected by 
the low-temperature tunneling microscope\cite{AY}. Thus the results of the bound state energies 
are valuable since they can be used to detect the symmetry of the excitonic insulators, especially for the materials
with qutie small splitting of singlet and triplet states.

 Another interesting finding is that magnetic impurity in triplet excitonic
insulator is
quite different from other cases. We can see the bound states caused by single
magnetic impurity in triplet exciton insulators are very close to the gap edge.
With the increasing of the impurity concentration an impurity band is expanding 
and the states are stretching to the Fermi level. Under this condition we could
expect that finite concentration of impurities should behave a weaker
pair breaking effect than in the other cases and we shall justify this
conclusion below.

\end {section}

\begin {section}{Finite concentration problems and pair-breaking effects}
To investigate the finite concentration problems we introduce the renormalized
Green Function. For spin singlet excitonic insulators:
\begin{equation}
G^{-1}(k,i\omega)=\left(\begin{array}{cc}
(\tilde{\omega}-\frac{\hbar^{2}}{2m}k^{2}+\frac{E_{g}}{2})\cdot I &
-\tilde{\Delta}_{S}\cdot I\\
-\tilde{\Delta}_{S}\cdot I &
(\tilde{\omega}+\frac{\hbar^{2}}{2m}k^{2}-\frac{E_{g}}{2})\cdot
I\end{array}\right)
\end{equation}
with the full Born approximation one can get the self-energy :
\begin{eqnarray*}
\sum(\omega)=n_{imp}\int\frac{d^{3}k}{(2\pi)^{3}}<VG(k,i\omega)V>_{I}
\end{eqnarray*}
where $<>_{I}$ means averaging all the impurity positions and $n_{imp}$ represents the density of impurities. Then we can
calculate the self-energy which is given by
\begin{eqnarray*}
\sum(\omega)=\left(\begin{array}{cc}
-\frac{\frac{1}{\tau_{s2\nu}}\tilde{\Delta}_{S}+\frac{1}{\tau_{s1\nu}}\tilde{
\omega}}{\sqrt{\Delta_{S}^{2}-\omega^{2}}}\cdot I &
-\frac{\frac{1}{\tau_{s1\nu}}\tilde{\Delta}_{S}+\frac{1}{\tau_{s2\nu}}\tilde{
\omega}}{\sqrt{\Delta_{S}^{2}-\omega^{2}}}\cdot I\\
-\frac{\frac{1}{\tau_{s1\nu}}\tilde{\Delta}_{S}+\frac{1}{\tau_{s2\nu}}\tilde{
\omega}}{\sqrt{\Delta_{S}^{2}-\omega^{2}}}\cdot I &
-\frac{\frac{1}{\tau_{s2\nu}}\tilde{\Delta}_{S}+\frac{1}{\tau_{s1\nu}}\tilde{
\omega}}{\sqrt{\Delta_{S}^{2}-\omega^{2}}}\cdot I\end{array}\right)
\end{eqnarray*}
and define the scattering times:
\begin{eqnarray*}
\frac{1}{\tau_{s1n}}&=&n_{imp}\pi N_{F}(J_{1n}^{2}+J_{2n}^{2})\\
\frac{1}{\tau_{s2n}}&=&2n_{imp}\pi N_{F}J_{1n}J_{2n}\\
\frac{1}{\tau_{s1m}}&=&n_{imp}\pi N_{F}(J_{1m}^{2}+J_{2m}^{2})S(S+1)\\
\frac{1}{\tau_{s2m}}&=&2n_{imp}\pi N_{F}J_{1m}J_{2m}S(S+1)
\end{eqnarray*}
As we know the renormalized Green Function can be written as:
\begin{eqnarray*}
G^{-1}(k,\omega)=G_{0}^{-1}(k,\omega)-\sum(\omega)
\end{eqnarray*}
then we have the self-consistent equations:
\begin{eqnarray}
\tilde{\omega}&=&\omega+\frac{\frac{1}{\tau_{s2\nu}}\tilde{\Delta}_{S}+\frac{1}{
\tau_{s1\nu}}\tilde{\omega}}{\sqrt{\Delta_{S}^{2}-\omega^{2}}}\\
\tilde{\Delta}_{S}&=&\Delta_{S}-\frac{\frac{1}{\tau_{s1\nu}}\tilde{\Delta}_{S}
+\frac{1}{\tau_{s2\nu}}\tilde{\omega}}{\sqrt{\Delta_{S}^{2}-\omega^{2}}}
\end{eqnarray}
Note here $\Delta_{S}$ represents the excitonic order parameter of singlet excitonic insulators in the presence of impurities. We consider the one-channel (inter-band or intra-band) impurity scattering
case, then the impurity effect only contributes to the renormalization of
frequency and the order parameter.
Making $\omega\rightarrow i\omega$, the order parameter in zero temperature and
the excionic gap can be got from:
\begin{eqnarray}
\Delta&=&VN_{F}\int_{0}^{\epsilon_{C}}d\omega\frac{\tilde{\Delta}}{\sqrt{\tilde{
\Delta}^{2}+\tilde{\omega}^{2}}}\\
\Omega&=&\Delta(1-\zeta^{2/3})^{3/2}
\end{eqnarray}
where $\epsilon_{C}$ is the cutoff and
$\zeta=[f(\tilde{\omega})-f(\tilde{\Delta})]/\Delta$. In the expression of
$\zeta$, $f(\tilde{\omega})$ and $f(\tilde{\Delta})$ can be well defined as the
factor of the renormalized frequency and order parameter if only one
impurity scattering channel opens ($J_{1\nu}=0$ or $J_{2\nu}=0$). They read
\begin{eqnarray*}
f(\tilde{\omega})&=&\frac{1}{\tau_{s1\nu}}\Big|_{J_{1\nu}\ or\ J_{2\nu}=0}\\
f(\tilde{\Delta})&=&-\frac{1}{\tau_{s1\nu}}\Big|_{J_{1\nu}\ or\ J_{2\nu}=0}
\end{eqnarray*}

The same analysis can be applied to the triplet case, in which the
renormalized Green-Function can be written as
\begin{eqnarray*}
G^{-1}(k,i\omega)=\left(\begin{array}{cc}
(\tilde{\omega}-\frac{\hbar^{2}}{2m}k^{2}+\frac{E_{g}}{2})\cdot I &
-\tilde{\Delta}_{T}\cdot\sigma\\
-\tilde{\Delta}_{T}\cdot\sigma &
(\tilde{\omega}+\frac{\hbar^{2}}{2m}k^{2}-\frac{E_{g}}{2})\cdot
I\end{array}\right)
\end{eqnarray*}

The self-consistent equations for one channel impurity scattering are:
\begin{eqnarray*}
\tilde{\omega}=\omega+(\frac{1}{\tau_{t1n}},\frac{1}{\tau_{t1m}})\frac{\tilde{
\omega}}{\sqrt{\tilde{\Delta}_{T}^{2}-\tilde{\omega}^{2}}}\\
\tilde{\Delta}_{T}=\Delta_{T}\mp(\frac{1}{\tau_{t1n}},\frac{1}{\tau_{t2m}})\frac
{\tilde{\Delta}_{T}}{\sqrt{\tilde{\Delta}_{_{T}}^{2}-\tilde{\omega}^{2}}}\\
\end{eqnarray*}
Note here the $(+,m)$ represents the magnetic case and $(-,n)$ represents the
nonmagnetic
case and $\Delta_T$ is the excitonic order parameter of triplet excitonic insulators in the presence of impurities. Then the scattering times read:
\begin{eqnarray*}
\frac{1}{\tau_{t1n}}&=&n_{imp}\pi N_{F}J_{in}^{2}\\
\frac{1}{\tau_{t1m}}&=&n_{imp}\pi N_{F}J_{im}^{2}S(S+1)\\
\frac{1}{\tau_{t2m}}&=&\frac{1}{3}n_{imp}\pi N_{F}J_{im}^{2}S(S+1)\\
\end{eqnarray*}
The numerical results for the order parameters and excitation gaps with the
impurity doping in singlet and triplet excitonic insulators are shown in Fig2.
\begin{figure}[htb]
\label{Fig2}\includegraphics[width=3.6in] {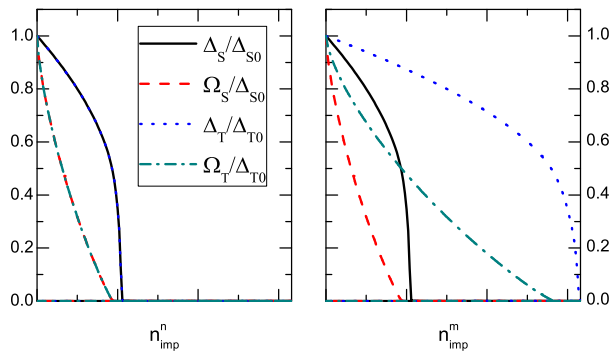}
\caption{The order parameters and the excitonic gaps via the
impurity concentration. The left figure is the nonmagnetic impurities in singlet
and triplet excitonic insulators and the right figure is for magnetic impurities
case. $\Delta_{S0}$ and $\Delta_{T0}$ represent the order parameter of singlet and triplet excitonic insulators in the absence of impurities.}
\end{figure}

From Fig2 we can see that both of magnetic and nonmagnetic impurities have the
pair breaking effects in excitonic insulators and there always exists a gapless
region between the gaped excitonic insulator phase and the normal phase. This
phase transition from gaped phase to gapless phase is of second order as pointed
out by J. Zittartz\cite{JZ}. However, the nonmagnetic impurities have the same
pair breaking effect to singlet and triplet excitonic insulators but magnetic
impurities behave differently in these two systems. A significant difference can
be seen from the right part in Fig2 in which the triplet excitonic insulators
are more robust with magnetic impurity doping. Notice that the pair
breaking effect coming from the self-consistent equations is decided by the
relative amplitude and sign between the factor of renormalized frequency and
order parameter in the self-energy.
Therefore, to compare the pair-breaking effect in quantization between different
cases, we define the pair breaking factor which can be written as
\begin{eqnarray*}
x=\frac{f(\tilde{\Delta})}{f(\tilde{\omega})}
\end{eqnarray*}

It is analogical to the notations in R. Balian et al 's work in p-wave
superconductors\cite{Balian}. According to Abrikosov-Gor'kov theory\cite{AG}, in
conventional s-wave superconductors, $x=-1$ stands for magnetic impurities and $x=1$
stands for nonmagnetic impurities.
Then we list all the pair breaking effects
in singlet and triplet excitonic insulators.
\begin{eqnarray}
x=&& \left\{\begin{array}{lcr}
-1\ \ \ \ \lambda_{S,n}\\
-1\ \ \ \ \lambda_{S,m}\\
-1\ \ \ \ \lambda_{T,n}\\
\frac{1}{3}\ \ \ \ \ \ \lambda_{T,m}
\end{array}\right. 
\end{eqnarray}
where $\lambda_{S,n}$, $\lambda_{S,m}$, $\lambda_{T,n}$, $\lambda_{T,m}$
represent the nonmagnetic impurity scattering and magnetic impurity
scattering in the intra-band/inter band channel in singlet and triplet
excitonic insulators, respectively. 

It is well known that the nonmagnetic impurities are not pair breakers
in s-wave conventional superconductors according to Anderson's theorem.
Note in singlet and triplet excitonic insulators nonmagnetic impurities
have the same pair-breaking effect. This is analogical to the magnetic
impurities in the s-wave superconductors in which the time reversal
symmetry is broken due to the spin-flip impurity scattering. The Kramer's
degeneracy is replaced by the particle-hole symmetry in the singlet
and triplet excitonic insulators. According
to our calculations triplet excitonic insulators are more stable with magnetic impurity
doping than singlet samples. $n_{T}/n_{S}=3$, where $n_{T}$ and
$n_{S}$ are the critical doping concentration of magnetic impurities
in triplet and singlet excitonic insulators, respectively. This result is in
accordance with the single impurity analysis. From Fig1 we can see that the
bound states energy is close to the gap edge for the magnetic impurity
in triplet excitonic insulators. And it needs larger impurity concentration in triplet than
in singlet excitonic insulators to fill the gap with states.
\end{section}

\begin {section}{Impurity effects and ferromagnetism}
In undoped CaB$_6$ the splitting of singlet and triplet excitonic phases is neglectable\cite{Gorkov},
then the $SU(2)\times SU(2)$ symmetry
is preserved and in this case the samples show no magnetic signals
since there is no polarization in the systems. All kinds of spin polarization
states have the same energy. Once the doping induce extra carriers
the degeneracy between singlet and triplet states can be lifted\cite{Rice,BAV}.
The
doped electrons are aligned in one direction to make the energy minimized.
Then the system have two order parameters $\Delta_{\uparrow}$ and
$\Delta_{\downarrow}$, which have the relationship with $\Delta_S$ and $\Delta_T$
in doped situation\cite{BAV}
\begin{eqnarray}
\Delta_{\downarrow}ln(\frac{\Delta_{S0}}{\Delta_{T0}})^{\frac{1}{2}}+\Delta_{
\uparrow}ln(\Delta_{S0}\Delta_{T0})^{\frac{1}{2}}&=&\Delta_{\uparrow}ln|\Delta_{
\uparrow}|  \nonumber\\
\Delta_{\uparrow}ln(\frac{\Delta_{S0}}{\Delta_{T0}})^{\frac{1}{2}}+\Delta_{
\downarrow}ln(\Delta_{S0}\Delta_{T0})^{\frac{1}{2}} \nonumber\\
=\Delta_{\downarrow}
ln(2n+\sqrt{4n^2+\Delta_{\downarrow}^2})
\end{eqnarray}
where $\Delta_{S0}$ and $\Delta_{T0}$ are the singlet and triplet excitonic order
parameters of undoped case in the absence of impurities, respectively.
$n$ is the concentration of the doping electrons. At zero doping at
$T=0$ we have $\Delta_{S0}=\Delta_{T0}=\Delta_0$ and in doped samples we have\cite{Rice,BAV}
\begin{eqnarray*}
&&\Delta_{\downarrow}=\sqrt{\Delta_{0}(\Delta_{0}-4n)} \ \ \ \ \
\rm{(0<n<\Delta/4)}\\
&&\Delta_{\uparrow}=\Delta_{0} \ \ \ \ \ \ \ \ \ \ \ \ \ \ \ \ \ \ \ \ \
\rm{(0<n<\Delta/2)}
\end{eqnarray*}
 Now we can see that
ferromagnetism can exist in the doping range of $[0,\Delta_{0}/2]$. In the doping level $[0,\Delta_{0}/4]$, the doped electrons are polarized (DP).
In the doping level $[\Delta_{0}/4,\Delta_{0}/2]$, all of the electrons
and holes are paired in one direction and system is in the fully polarized phase (FP). When $n>\Delta_{0}/2$, the excitonic
state is no longer favored and system undergoes a first order transition into the normal metallic
state. This phase transition is different with the excitonic insulator-metal
phase transition we discussed above. The former one is caused by carrier doping
and becomes a normal metal after the transition. The latter one is caused by the
impurity doping and undergoes a transition into the impure metal phase. Besides,
for the impurity induced excitonic insulator-metal phase transition, generally
there is a second order phase transition into a gapless region before the
excitonic insulator-metal phase transition.

In bulk materials the La doping would induce impurities or disorders
inevitably and in this case both the carrier doing and impurity effect must be
considered. For CaB$_6$, since there is no exact experimental data of
$\Delta_{0}$ and if
we consider a general excitonic gap of order $\backsim600K$\cite{Y}, the
ferromagnetism range via doping should be $0-3\%$ and this doping
range is much larger than the experimental results $0-1\%$. If
the excitonic insulators scenario is responsible for the physics in
CaB$_6$, one possible reason for the narrow range of ferromagnetism
is the anisotropy of the band structure\cite{Rice,Gorkov}. However
this is not enough and we must involve the pair breaking effects caused
by impurity or disorder. A direct evidence for this comes from the tunnelling
experiments\cite{BK}, in which it is surprising to find that there exists
ferromagnetism
in 99.9\% pure boron of CaB$_6$ within a small range of carrier
doping while there are no magnetic signals in 99.9999\% samples. For high purity samples
the exciton gap may be too large compared to the doping concentration and the splitting between $\Delta_{\downarrow}$ and $\Delta_{\uparrow}$ is
neglectable.
Therefore the ferromagnetism can not be found. More disorders or impurities in
the samples suppress the excitonic order parameters and make it comparable to the doping level.
Then the impurity effect results in two changes: a more significant splitting of
$\Delta_{\downarrow}$ and $\Delta_{\uparrow}$ and a narrower doping range of
ferromagnetism.

To express our physics in detail we carry out the numerical work by replacing
the $\Delta$ in (1) with
$\Delta_{S}\cdot I+\Delta_{T}\cdot\sigma$. For simplification we choose
the triplet exciton along the z direction. Then with the same full
Born approximation we have the self-consistent equations with nonmagnetic
impurity scattering:
\begin{eqnarray}
\tilde{\omega}_{1}&=&\omega+\frac{\tilde{\omega}_{1}\frac{1}{\tau_{t1n}}}{\sqrt{
(\tilde{\Delta}_{S}+\tilde{\Delta}_{T})^{2}-\tilde{\omega}_{1}^2}} \nonumber\\
\tilde{\omega}_{2}&=&\omega+\frac{\tilde{\omega}_{2}\frac{1}{\tau_{t1n}}}{\sqrt{
(\tilde{\Delta}_{T}-\tilde{\Delta}_{S})^{2}-\tilde{\omega}_{2}^2}} \nonumber\\
\tilde{\Delta}_{T}+\tilde{\Delta}_{S}&=&(\Delta_{T}+\Delta_{S})-\frac{(\tilde{
\Delta}_{T}+\tilde{\Delta}_{S})\frac{1}{\tau_{t1n}}}{\sqrt{(\tilde{\Delta}_{T}
+\tilde{\Delta}_{S})^{2}-\tilde{\omega}_{1}^2}} \nonumber\\
\tilde{\Delta}_{T}-\tilde{\Delta}_{S}&=&(\Delta_{T}-\Delta_{S})-\frac{(\tilde{
\Delta}_{T}-\tilde{\Delta}_{S})\frac{1}{\tau_{t1n}}}{\sqrt{(\tilde{\Delta}_{T}
-\tilde{\Delta}_{S})^{2}-\tilde{\omega}_{2}^2}} \nonumber\\
\end{eqnarray}
the self-consistent equations with magnetic impurity scattering:
\begin{eqnarray}
\tilde{\omega}_{1}=\omega&+&\frac{\tilde{\omega}_{1}\frac{1}{\tau_{t2m}}}{\sqrt{
(\tilde{\Delta}_{T}+\tilde{\Delta}_{S})^{2}-\tilde{\omega}_{1}^2}}+\frac{\tilde{
\omega}_{2}\frac{2}{\tau_{t2m}}}{\sqrt{(\tilde{\Delta}_{T}-\tilde{\Delta}_{S})^{
2}-\tilde{\omega}_{2}^2}} \nonumber\\
\tilde{\omega}_{2}=\omega&+&\frac{\tilde{\omega}_{1}\frac{2}{\tau_{t2m}}}{\sqrt{
(\tilde{\Delta}_{T}+\tilde{\Delta}_{S})^{2}-\tilde{\omega}_{1}^2}}+\frac{\tilde{
\omega}_{2}\frac{1}{\tau_{t2m}}}{\sqrt{(\tilde{\Delta}_{T}-\tilde{\Delta}_{S})^{
2}-\tilde{\omega}_{2}^2}} \nonumber\\
\tilde{\Delta}_{T}+\tilde{\Delta}_{S}&=&(\Delta_{T}+\Delta_{S})-\frac{(\tilde{
\Delta}_{T}+\tilde{\Delta}_{S})\frac{1}{\tau_{t2m}}}{\sqrt{(\tilde{\Delta}_{T}
+\tilde{\Delta}_{S})^{2}-\tilde{\omega}_{1}^2}} \nonumber\\
&&+\frac{(\tilde{\Delta}_{T}-\tilde{\Delta}_{S})\frac{2}{\tau_{t2m}}}{\sqrt{
(\tilde{\Delta}_{T}-\tilde{\Delta}_{S})^{2}-\tilde{\omega}_{2}^2}} \nonumber\\
\tilde{\Delta}_{T}-\tilde{\Delta}_{S}&=&(\Delta_{T}-\Delta_{S})+\frac{(\tilde{
\Delta}_{T}+\tilde{\Delta}_{S})\frac{2}{\tau_{t2m}}}{\sqrt{(\tilde{\Delta}_{T}
+\tilde{\Delta}_{S})^{2}-\tilde{\omega}_{1}^2}} \nonumber\\
&&-\frac{(\tilde{\Delta}_{T}-\tilde{\Delta}_{S})\frac{1}{\tau_{t2m}}}{\sqrt{
(\tilde{\Delta}_{T}-\tilde{\Delta}_{S})^{2}-\tilde{\omega}_{2}^2}} \nonumber\\
\end{eqnarray}
Taking both the impurity effects and carrier doping into considerations, we have to solve the Equs.(6), (9) and (10) for nonmagnetic impurities and Equs. (6), (9) and (11) for magnetic impurities. Now the
ferromagnetic range and the fully polarized range should be $0-n_1$ and $n_2-n_1$,
where the critical charged impurity concentration $n_1$ and $n_2$ satisfy
\begin{eqnarray*}
\Delta_{\uparrow}(n^C_{imp}=n_1)=0\\
\Delta_{\downarrow}(n^C_{imp}=n_2)=0
\end{eqnarray*}
Note here the impurity concentration $n_{imp}$ and density of doping electrons should be replaced with charged impurity concentration $n^C_{imp}$.
Our numerical results are shown in Fig3. 
\begin{figure}[htb]
\label{Fig3}\includegraphics[width=3.6in] {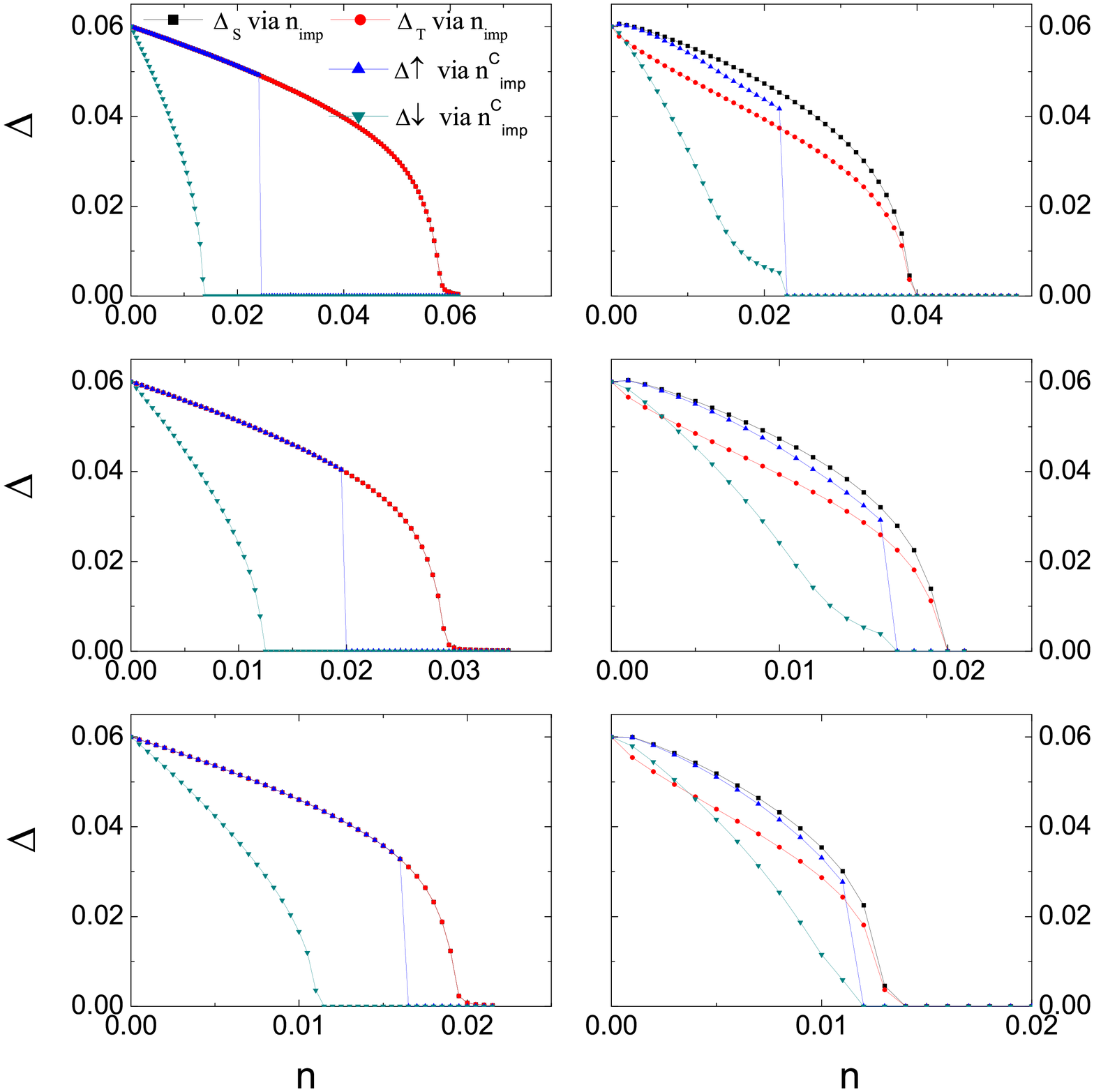}
\caption{Order parameters with doping concentration. Note in the
figure $n^{C}_{imp}$ and $n_{imp}$ represent the charged and non-charged impurities
concentration, respectively. From top to below:
$\frac{1}{\tau_{t1n}n},\frac{1}{\tau_{t2m}n}=0.5$;
$\frac{1}{\tau_{t1n}n},\frac{1}{\tau_{t2m}n}=1$ and  $\frac{1}{\tau_{t1n}n},\frac{1}{\tau_{t2m}n}=1.5$.}
\end{figure}
\begin{figure}[htb]
\label{Fig4}\includegraphics[width=3.2in] {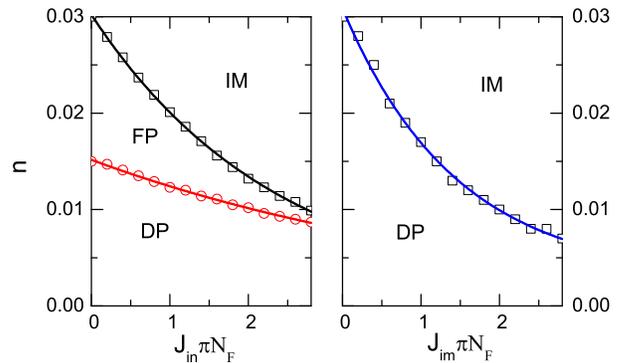}
\caption{Phase diagram in the doping concentration-impurity
scattering plane. In the figure IM  represent the impure metal phase. FP and DP
represent the fully polarized excitonic insulator phase for all the
electron-hole pairs and all doped carriers, respectively. Note the weak
ferromagnetic region includes FP and DP region.}
\end{figure}

In Fig3 we note that in the nonmagnetic
impurity scattering case, the pair breaking effect is the same for
singlet and triplet excitons. But in the magnetic impurity scattering
case, the order parameter of singlet and triplet excitons splits
with the increasing of the impurity concentration. Both two kinds of order
parameters disappear at the same doping concentration at which the
excitonic insulator becomes an impure metal. In the situation here we do not
consider the gapless excitonic insulator since this regain is quite
small($<0.05\Delta_0$). From top to below in Fig3 we can also see that the ferromagnetic region $0-n_1$ is growing narrower with the increasing of the impurity scattering. To express that clearer we plot the phase diagram in $J-n$ plane in Fig4. It is
very clear that the nonmagnetic and magnetic impurities would suppress the
ferromagnetic range via doping. In magnetic impurity case, the fully polarized
range is destroyed by the impurity scattering in all doping levels. For pure
carriers doping the ferromagnetic range should be $0-3\%$ and this region are
suppressed to $0-1\%$ with the impurity scattering $J_{in}\pi N_F\simeq2.8$ for nonmagnetic case and $J_{im}\pi N_F\simeq2.0$ for magnetic case, where $i=1,2$.
And this result is in accordance with the experimental data. It is worthy to point out that here we considered the charged impurity case which means the density of doped electrons and impurities have the same value $n=n_{imp}=n^{C}_{imp}$. In CaB$_6$, the charged impurities are induced by La doping. If there are additional impurities/disorders which are not accompanied with electrons doping, which means $n_{imp}>n$, they can strengthen the pair-breaking effects and these extra impurity scattering terms can be absorbed into the factor of $J_{in}$ or $J_{im}$ directly.

\end {section}

\begin {section}{Discussion and summary}
We investigated the nonmagnetic and magnetic impurity effects in singlet
and triplet excitonic insulators in the inter-band and intra-band channels.
In single impurity problems, the bound states showed different compositions in singlet and triplet excitonic insulators, which can be used to detect the symmetry of excitonic insulators in experiments.  In finite concentration problems, all kinds of impurities showed pair-breaking effects. In the presence of impurities, the excitonic insulator scenario for ferromagnetism in
CaB$_6$ was modified . The doped excitonic
insulators can be characterized in three classes by different kinds of doping:

(1) Electrons doping but no impurity effects induced: Ferromagnetism emerges
with the electrons doping and there are two ferromagnetic phases with doping,
the polarized phase for the doped electrons and the full polarized phase for all
the excitonic pairs.

(2) Impurity effects induced but no electrons brought in:  i. Nonmagnetic
impurities in one channel show same pair breaking effects to singlet and triplet
excitonic insulators. ii. Magnetic impurities in one channel show weaker pair
breaking effect to triplet excitonic insulators than singlet ones. 

(3) Both electrons and impurity effect were induced by doping: Nonmagnetic and
magnetic impurity effects suppressed the ferromagnetic region and this provides
an natural explaination for the quite narrow region for ferromagnetism in the
experiments of CaB$_6$.  In the magnetic impurity case, the fully polarized phase for
all the electron-hole pairs was destroyed in all doping levels.

The model we discussed here is based on two nested bands which have the
same effective masses. Once apart from this situation, the two bands have
different effective masses, $m_{a}\neq m_{b}$. In this case the impurities are
still pair breakers because they can be seen as oppositely charged pairs to an
excitoni\cite{JZ}. This may make the quantitative calculations much more
difficult but should show the same conclusions as ours qualitatively. The more
anisotropy of band structure will obviously reduce the total magnetism and may
lead to the well-known Larkin-Ovchinnikov-Fulde-Ferrerll state\cite{Rice,Gorkov}
which is not considered in this work.

We would like to thank Chin-Sen Ting, Xiaoming Cai, Zi Cai and Xiaoling Cui for helpful discussions. This work was financially supported
by NSFC, CAS and 973-project of MOST of China.
\end {section}

\end{document}